\documentclass[final,8pt,5p,twocolumn]{elsarticle}

\usepackage{amssymb}

\usepackage[final]{changes}

\usepackage{forest}

\usepackage{ulem}

\usepackage{xcolor}

\usepackage{tabularx,ragged2e}
\newcommand{\HY}{\hyphenpenalty=25\exhyphenpenalty=25} 
\newcolumntype{Z}{>{\HY\RaggedRight\arraybackslash\hspace{0pt}}X} 

\usepackage[hyphens]{url}
\usepackage{hyperref} 
\hypersetup{breaklinks=true}
\usepackage{caption}
\usepackage{booktabs}
\usepackage{multirow}
\usepackage{tabularx}	
\usepackage{color,colortbl}
\usepackage{subcaption}
\usepackage{graphicx}  

\usepackage[smartEllipses]{markdown}

\makeatletter
\newcommand{\ion}[2]{%
  #1$\;$%
  \if b\expandafter\@car\f@series\relax\@nil
    \begingroup 
      \sbox0{\rmfamily\mdseries\textsc{v}}%
      \resizebox{!}{\ht0}{\rmfamily\@Roman{#2}}%
    \endgroup
  \else
    \textsc{\rmfamily\@roman{#2}}%
  \fi
}
\makeatother

\makeatletter
\def\@author#1{\g@addto@macro\elsauthors{\normalsize%
    \def\baselinestretch{1}%
    \upshape\authorsep#1\unskip\textsuperscript{%
      \ifx\@fnmark\@empty\else\unskip\sep\@fnmark\let\sep=,\fi
      \ifx\@corref\@empty\else\unskip\sep\@corref\let\sep=,\fi
      }%
    \def\authorsep{\unskip,\space}%
    \global\let\@fnmark\@empty
    \global\let\@corref\@empty  
    \global\let\sep\@empty}%
    \@eadauthor={#1}
}
\makeatother

\journal{Journal of Systems and Software}

\begin{document}

\begin{frontmatter}

\title{From Paradigm Shift to Audit Rift: Empirical Analysis and Validation of Security Audit Methodologies for Asynchronous Smart Contract Systems}


\affiliation[Sk]{organization={Skolkovo Institute of Science and Technology},
            city={Moscow},
            country={Russia}}


\affiliation[PT]{organization={Positive Technologies},
            city={Moscow}, country={Russia}}

\affiliation[MSU]{organization={Lomonosov Moscow State University},
            city={Moscow}, country={Russia}}

\affiliation[MIPT]{organization={Moscow Institute of Physics and Technology},
            city={Moscow},
            country={Russia}}

\affiliation[IIT]{organization={Indian Institute of Technology},
            city={Delhi},
            country={India}}

\author[Sk]{Yury Yanovich}
\ead{{Corresponding author*}y.yanovich@skoltech.ru}

\author[PT]{Sergey Sobolev}
\ead{ssobolev@ptsecurity.com}

\author[Sk]{Yash Madhwal}
\ead{yash.madhwal@skoltech.ru}

\author[PT,MSU]{Kirill Ziborov}
\ead{kziborov@ptsecurity.com}

\author[MIPT]{Vladimir Gorgadze}
\ead{gorgadze@gmail.com}

\author[MIPT]{Victoria Kovalevskaya}
\ead{vvkovalevskaya@phystech.edu} 

\author[MIPT]{Elizaveta Smirnova}
\ead{smirnova.elizaveta.s@phystech.edu}

\author[MIPT]{Matvey Mishuris}
\ead{mishuris.ma@phystech.edu}

\author[IIT]{Subodh Sharma}
\ead{svs@iitd.ac.in}  

\begin{abstract}
    The Open Network (TON) is a high-performance blockchain platform designed for scalability and efficiency, leveraging an asynchronous execution model and a multi-layered architecture. While TON's design offers significant advantages, it also introduces unique challenges for smart contract development and security. This paper introduces a comprehensive audit checklist for TON smart contracts, based on an empirical analysis of 34 professional audit reports containing 233 real-world vulnerabilities. The checklist addresses TON-specific challenges, such as asynchronous message handling, and provides actionable insights for developers and auditors. We also present detailed case studies of vulnerabilities in TON smart contracts, highlighting their implications and offering lessons learned. To validate practical utility, we conducted a practitioner survey (n=11 complete responses), confirming the checklist's value alongside automated tools. By adopting this checklist, developers and auditors can systematically identify and mitigate vulnerabilities, enhancing the security and reliability of TON-based projects. Our work bridges the gap between Ethereum's mature audit methodologies and the emerging needs of the TON ecosystem, fostering a more secure and robust blockchain environment.
\end{abstract}

\begin{keyword}
TON \sep The Open Network \sep Smart Contracts \sep Security Audits \sep Asynchronous Execution \sep FunC
\end{keyword}

\end{frontmatter}

\section{Introduction}
\label{section:Introduction}

The Open Network (TON) is a next-generation blockchain platform designed to achieve unparalleled scalability and efficiency through its innovative multi-layered architecture \cite{Durov2019}. Originally conceptualized by Telegram, TON has transitioned into an independent open-source project following regulatory challenges \cite{Berger2019}. The platform supports a wide range of decentralized applications (dApps), smart contracts, and micropayment services, powered by its native cryptocurrency, Toncoin, which facilitates transaction processing and ensures network security~\cite{TON2025}.

TON's architecture is built around a masterchain that coordinates multiple workchains and shardchains, enabling high throughput and scalability. Whilst the original design envisioned up to $2^{32}$ workchains and $2^{60}$ shardchains, the current implementation operates with two workchains, each containing a single shard \cite{TON2025a}. This streamlined design strikes a balance between scalability and operational efficiency while maintaining flexibility to expand as the network grows.

At the core of TON's smart contract execution is the TON Virtual Machine (TVM), a stack-based virtual machine optimized for high-performance and deterministic execution \cite{Durov2020}. TVM provides developers with a rich set of primitives for stack manipulation, arithmetic operations, control flow, and data structure management, enabling the implementation of complex logic with efficiency. However, TON's asynchronous execution model, which processes transactions and smart contract interactions concurrently through a message-passing mechanism, introduces unique challenges. Unlike synchronous blockchains such as Ethereum, where transactions are processed sequentially, TON's asynchronous model enhances scalability and reduces latency but requires careful design to mitigate potential security risks, including race conditions and message replay attacks.

Despite the increasing adoption of the TON blockchain, the field of smart contract security on TON remains underdeveloped compared to Ethereum \cite{Buterin2014}. Existing research on smart contract vulnerabilities predominantly focuses on the Solidity programming language and the Ethereum blockchain \cite{Atzei2017,Chen2022a,Zhang2020b,Soud2024}. While these studies provide a solid foundation for understanding vulnerabilities in Solidity-based smart contracts, they also highlight the prevalence of language-specific and network-specific coding issues. This underscores the need for tailored methodologies in other ecosystems, such as TON, which employs an asynchronous transaction execution model, the FunC programming language, and a sharding mechanism for scalability.

This paper addresses this critical need by presenting a detailed checklist of potential vulnerabilities in TON smart contracts and proposing a structured audit workflow. Our primary contributions include: 
\begin{enumerate}
    \item a comprehensive audit checklist tailored to TON smart contracts
    \item an analysis of 34 professional audit reports to identify common vulnerabilities
    \item  detailed case studies illustrating the implications of these vulnerabilities
    \item empirical validation of the checklist via a practitioner survey (n=11).
\end{enumerate}
The checklist is derived from an in-depth analysis of professional audit reports for TON-based projects, encompassing a wide range of vulnerabilities, including logical errors, access control issues, and state inconsistencies. By adopting this checklist, developers and auditors can systematically identify and mitigate vulnerabilities, thereby enhancing the security and reliability of TON-based projects. Our work bridges the gap between Ethereum's mature audit methodologies and the emerging needs of the TON ecosystem, contributing to a more secure and robust blockchain environment.

The remainder of this paper is organized as follows. Section \ref{section:Background} provides an overview of the TON architecture and its implications for smart contract development. Section \ref{section:RelatedWork} discusses related work. Section \ref{section:AuditsReview} outlines the methodology used to collect and analyze audit reports for TON smart contracts. Section \ref{section:Checklist} introduces the comprehensive audit checklist for TON smart contracts, organized into key categories such as contract design, asynchronous execution, and common vulnerabilities. Section \ref{section:VulnerabilitiesCaseStudy} presents detailed case studies of vulnerabilities in TON smart contracts, emphasizing their implications and lessons learned. Section \ref{section:Discussion} discusses the broader implications of our findings for the TON ecosystem. Finally, Section \ref{section:Conclusion} concludes the paper with recommendations for future research and development.

\section{Background: Architecture of TON}  
\label{section:Background}   

\subsection{TON Virtual Machine (TVM)} 

The main purpose of the Telegram Open Network Virtual Machine (TVM) \cite{Durov2020} is to run smart contract code on the TON Blockchain. It is designed to manage everything needed to process incoming messages and stored data while also creating new messages and making changes to the stored data. The state of TVM is defined by 5 key properties, which are described in more detail below: \textbf{Stack}, \textbf{Control Registers}, \textbf{Current Continuation}, \textbf{Current Codepage}, and \textbf{Gas Limits}.

\subsubsection{Stack and Data Management}  
TVM is a stack-based virtual machine, akin to the one used in Ethereum. As a stack-based machine, it operates on Last-In-First-Out principle (LIFO), where data are stored and retrieved from the top of the stack. Most operations and functions take their input from the top of the stack and replace them with the result. The stack can hold 7 types of variables: \textbf{Integer}, \textbf{Cell} (stores up to 1023 bits of data and references to up to 4 other cells), \textbf{Null}, \textbf{Tuple} (an ordered collection of up to 255 items), \textbf{Slice} (a read-only view of part of a cell), \textbf{Builder} (used to pack data into new cells), and \textbf{Continuation} (allows a cell to act as a set of TVM instructions). TVM’s support for advanced data structures, such as tuples and dictionaries, ensures efficient storage and retrieval of key-value pairs, which is critical for managing complex smart contract state.

\subsubsection{Control Registers}  
Some values that are rarely changed but frequently used across functions are better stored in special registers rather than the stack. This avoids excessive stack reordering. To address this, TVM includes up to 16 control registers, labeled \textbf{c0} to \textbf{c15}. These registers are particularly useful for managing control flow and temporary data, as they reduce the need for repetitive stack manipulation operations like \texttt{PUSH}, \texttt{POP}, and \texttt{XCHG}.

\subsubsection{Current Continuation}  
The \textbf{Current Continuation} represents the sequence of instructions to be executed after the current operation is completed. Continuation primitives in TVM, such as dynamic jumps and subroutine calls, enable flexible control flow management. This is essential for implementing complex logic, including loops and exception handling, which are supported by opcodes like \texttt{REPEAT}, \texttt{WHILE}, and \texttt{AGAIN}.

\subsubsection{Current Codepage}  
The \textbf{Current Codepage} is a value that determines how the next TVM opcode will be decoded. This feature allows TVM to adapt to different instruction sets or encoding schemes, enhancing its versatility for various smart contract use cases.

\subsubsection{Gas Limits}  
\textbf{Gas Limits} include four values: the current gas limit, the maximum gas limit, the remaining gas, and the gas credit. Efficient gas management is critical for ensuring that smart contracts are executed within resource constraints, avoiding excessive computational costs. TVM’s gas-related primitives enable developers to optimize smart contract execution by monitoring and controlling gas consumption.

\subsubsection{TVM Instructions (Primitives)}  
TVM instructions, often referred to as primitives or built-in operations, are the smallest, indivisible actions that TVM can perform as part of its code. These instructions are grouped into categories based on the types of values they handle (e.g., integers, cells, etc.). For example, TVM supports arithmetic operations like addition, subtraction, multiplication, and division, as well as bitwise shifts and logical operations. It also includes specialized primitives for blockchain-specific tasks, such as message handling and cryptographic operations, which enable smart contracts to interact seamlessly with the TON Blockchain environment.

By combining these features, TVM provides a robust and efficient execution engine for smart contracts, making it a critical component of the TON ecosystem.
Understanding these elements is essential for developing a comprehensive audit checklist that ensures the security, efficiency, and correctness of TON smart contracts.

\subsection{Asynchronous Transaction Execution}  
TON employs an asynchronous execution model, where transactions and smart contract interactions are processed concurrently through a message-passing mechanism. Unlike synchronous blockchains like Ethereum, where transactions are processed sequentially, TON’s asynchronous model allows for higher scalability and reduced latency. Messages between smart contracts are queued and processed independently, enabling parallel execution.  

This model introduces unique challenges for smart contract developers, such as managing message dependencies and ensuring atomicity in asynchronous interactions. While asynchronous execution enhances scalability, it also requires careful design to mitigate potential security risks, such as race conditions and message replay attacks.  

\subsection{Bridging TON Asynchronous Model and Solidity Synchronous Model}  
In contrast to TON’s asynchronous model, Solidity--the programming language for Ethereum--operates synchronously~\cite{Buterin2014}. In Ethereum, transactions are executed sequentially within a block, ensuring deterministic execution but limiting scalability. Conversely, TON’s asynchronous model enables parallel processing of transactions, significantly improving throughput while also presenting challenges for smart contract designers and developers. 

However, this difference introduces a paradigm shift for developers transitioning from Solidity to TON. In Solidity, developers can rely on the sequential execution of transactions, simplifying state management. In TON, developers must account for the non-deterministic nature of message processing, requiring robust design patterns to handle message dependencies and ensure consistency.  

\subsection{Implications for Smart Contract Development}  
The asynchronous execution model of TON provides significant benefits, including improved scalability and reduced transaction latency. However, it also introduces complexities in smart contract development, as developers must account for the non-deterministic nature of message processing. Proper design patterns and security audits are essential to ensure the reliability and safety of smart contracts on TON. 

\section{Related Work}
\label{section:RelatedWork}

\subsection{Smart Contract Vulnerabilities beyond TON}  
Currently, researchers of smart contract vulnerabilities predominantly focus on the Solidity language and the Ethereum blockchain. Atzei et al.~\cite{Atzei2017} provided the first comprehensive analysis, classifying 12 vulnerabilities into three levels: Solidity, EVM, and blockchain. Subsequent studies have further refined this classification. Chen et al.~\cite{Chen2022a} used StackExchange posts to identify vulnerabilities and categorize them according to security levels, while Zhang et al.~\cite{Zhang2020b} collected bugs from academic literature and open-source projects, offering a categorization based on the source of errors. Soud et al.~\cite{Soud2024} integrated existing classification schemes, identifying 47 unique patterns from 2143 vulnerabilities and grouping them into 11 categories.  

While these studies provide a robust foundation for understanding vulnerabilities in Solidity smart contracts, they also highlight the dominance of language-specific and network-specific coding issues. This underscores the need for tailored methodologies in other ecosystems, such as the TON blockchain, which employs the FunC programming language and features a sharding mechanism for scalability. Our work addresses this gap by proposing a comprehensive audit checklist tailored to TON smart contracts.  

\subsection{Smart Contracts Audit Tools}  
Modern audit methodologies combine manual analysis with automated tools for vulnerability detection. He et al.~\cite{He2020} summarized existing Ethereum smart contract security audit methods and compared several mainstream tools. Khan and Namin~\cite{Khan2020, Khan2024} classified the vulnerabilities and mapped them to 41 detection tools. Current approaches include static analysis, dynamic analysis, and formal verification.  

Static analysis tools, such as Slither~\cite{Feist2019}, examine the Solidity source code for security vulnerabilities by converting its abstract syntax tree into an intermediate representation. MythX~\cite{Consensys2025}, a platform developed by ConsenSys, combines static analysis (via Mythril~\cite{Consensys2025a}) with dynamic analysis (via Harvey~\cite{Wustholz2020}). Mythril employs symbolic execution and the Z3 SMT solver~\cite{deMoura2008}, while Harvey is a fuzzing tool that generates random inputs to test for unexpected behavior. Other fuzzers, such as Echidna~\cite{Grieco2020} and sFuzz~\cite{Nguyen2020a}, are also widely used.  

Kalra et al.~\cite{Kalra2018} introduced ZEUS, a formal verification framework for analyzing the safety of smart contracts. ZEUS leverages abstract interpretation, symbolic model checking, and constrained Horn clauses to verify contracts for correctness and fairness. Evaluated on over 22,400 Ethereum and Hyperledger Fabric smart contracts, ZEUS found that approximately 94.6\% contracts were vulnerable to issues such as reentrancy, unchecked send, and integer overflow/underflow. While ZEUS demonstrates the effectiveness of formal verification in identifying vulnerabilities, its focus remains on Ethereum and Fabric, leaving a gap in methodologies for other ecosystems like TON.  

Other formal verification tools, including KEVM~\cite{Hildenbrandt2018}, Certora Prover~\cite{Certora2025}, and Concert~\cite{Annenkov2020}, provide rigorous proofs of correctness for smart contracts. Certora Prover, in particular, has been extended to support multiple blockchain platforms, including Solana and Stellar, demonstrating the potential for cross-chain applicability. However, these tools have yet to be adapted for TON, presenting an opportunity for future research.  

Recent advancements in smart contract auditing, such as the use of machine learning models like Distil-BERT for vulnerability detection, highlight the limitations of traditional static analysis in identifying complex vulnerabilities. For example, SmartLLM leverages generative AI to detect vulnerabilities in Ethereum smart contracts, providing intuitive justifications for its findings~\cite{Kevin2025}. Similarly, Ma et al. propose combining fine-tuning and LLM-based agents to enhance the accuracy and interpretability of smart contract audits~\cite{Ma2025}. Whilst these techniques have been primarily applied to Ethereum, their potential applicability to TON remains unexplored. Additionally, formal verification tools like Certora Prover, which support multiple blockchain platforms, could be adapted for TON, though this requires further investigation.  

\subsection{TON Smart Contracts Security Analysis}  
The field of smart contract security on the TON blockchain is underexplored compared to Ethereum. Song et al.~\cite{Song2025} identified 8 types of vulnerabilities in TON smart contracts and introduced TONScanner, a static analysis framework for detecting them. Ton Symbolic Analyzer (TSA)~\cite{Espirito2025}, a newer industrial tool, operates at the bytecode level and uses symbolic virtual machines and SMT solvers. These tools represent initial steps toward addressing TON-specific vulnerabilities, but they lack the comprehensive coverage of their Ethereum counterparts.  

Our work builds on these foundations by presenting a detailed checklist of potential vulnerabilities in TON smart contracts and proposing a structured audit workflow. This workflow addresses the unique challenges posed by the TON's architecture, such as its use of FunC and its focus on high throughput and scalability. By doing so, we aim to bridge the gap between Ethereum's mature audit methodologies and the emerging needs of the TON ecosystem.

\section{Smart Contract Security Audits Review}
\label{section:AuditsReview}
To ensure a comprehensive and empirical collection of vulnerability data, we employed a structured methodology focused on industry artifacts rather than academic literature. We identified security firms and audit agencies that have publicly published audit reports for TON-based projects, including BugBlow, TonBit, ScaleBit, Beosin, Quantstamp, and HashEx Blockchain Security. Additionally, we explored public repositories (e.g., GitHub, GitLab) and community-driven platforms (e.g., TON-specific forums, Telegram groups, Discord channels) for shared reports. 

The data collection strategy involved systematic searches using keywords such as \textit{TON}, \textit{The Open Network}, \textit{TVM}, \textit{smart contract}, \textit{audit}, and \textit{security report}. Inclusion criteria focused on relevance to TON smart contracts, recency (2023--2025), and public accessibility. This approach yielded a robust dataset of 34 reports \cite{Tonbit2023TonUp,Quantstamp2023TONLocker,Sedov2023Hipo,ScaleBit2023Hipo, Beosin2024Aqua,Beosin2024InterBridge,Beosin2024Onton,Beosin2024TONCO,Beosin2024Tonny, BugBlow2024Aqua,BugBlow2024CryptoBillions, TonBit2024BeaverLand,TonBit2024Bool,TonBit2024MiniTon,TonBit2024OneClick,Tonbit2024SecondLive,TonBit2024Catizen,TonBit2024CatizenJetton,Tonbit2024Thunder,TonBit2024TOMPUMP,Tonbit2024TonStaking,Tonbit2024Trapdoor,Tonbit2024TRC404,Tonbit2024UTonic, Quantstamp2024FDUSD,Quantstamp2024Storm,HashEx2024Grishmans,Quantstamp2024RhinoFi,Quantstamp2024Evaa, TonTech2024Hipo,ProgramCrafter2024Hipo,Chainsulting2024TONMS,Softstackio2024XTON, BugBlow2025Boxing} covering 29 projects, conducted by 11 audit companies, groups, or individuals. These reports covered projects in FunC (28 reports) and Tact (6 reports). The projects spanned diverse categories, including DeFi protocols, NFT platforms, gaming projects, and staking protocols. 

\subsection{Vulnerability Distribution}  
The distribution of reports varied by company, with TonBit contributing 14 reports, Beosin and Quantstamp each contributing 5, and BugBlow contributing 3. The number of vulnerabilities per report ranged up to 32, with one report declaring no vulnerabilities. In total, 233 vulnerabilities were reported across the remaining reports (see Table \ref{tab:VulnerabilityByLevel}). The "undetermined" security level refers to vulnerabilities flagged by the audit team during the process.  

\begin{table}[h!]  
\caption{Vulnerability Count by Security Level}  
\label{tab:VulnerabilityByLevel}  
\centering  
\begin{tabular}{lc}  
\toprule  
\textbf{Security Level} & \textbf{Vulnerability Count} \\  
\midrule  
Critical & 11 \\  
Major & 35 \\ 
Medium & 53 \\  
Low & 71 \\ 
Informational & 49 \\  
Undetermined & 14 \\  
\midrule  
\textbf{Total} & 233 \\  
\bottomrule  
\end{tabular}  
\end{table}  

\subsection{Common Vulnerabilities}  
Common vulnerabilities included logical errors (e.g., incorrect state transitions, integer overflow/underflow), access control issues (e.g., unauthorized function calls, privilege escalation), and state inconsistency (e.g., delayed state updates, race conditions). Reentrancy vulnerabilities, though less frequent than in synchronous blockchains like Ethereum, were still present. Message ordering and timing attacks, particularly in DeFi and gaming applications, were also identified.  

Logical errors and access control issues were the most frequent, appearing in approximately 70\% of the reports. Reentrancy and state inconsistency, though less common, had a high potential impact. Message ordering and timing attacks were rare but critical, underscoring the unique challenges of TON's asynchronous execution model.  

\subsection{Implications for Audit Practices}  
Our analysis of these audit reports highlights several key insights for improving audit practices in the TON ecosystem:  

\begin{itemize}  
\item \textbf{Logical Errors}: These were the most prevalent vulnerabilities, emphasizing the need for rigorous testing and validation of contract logic.  
\item \textbf{Access Control Issues}: Robust authorization checks are essential to prevent unauthorized function calls and privilege escalation.  
\item \textbf{Asynchronous Execution Challenges}: Delayed state updates and race conditions are unique to TON's asynchronous model, and require careful design and auditing.  
\item \textbf{Gas Control}: Improper gas handling can lead to state inconsistencies and financial losses, underscoring the importance of accurate gas calculations.  
\end{itemize}  

These insights formed the development of our comprehensive audit checklist, which addresses these vulnerabilities and provides actionable recommendations for developers and auditors.

\section{Checklist for Auditing TON Smart Contracts}
\label{section:Checklist}

In this section, we introduce our comprehensive checklist for auditing TON smart contracts, which is publicly available on GitHub \cite{PositiveSecurity2025}.
The checklist is derived from publicly available audit reports and our internal auditing practices. It serves as a heuristic tool designed to minimize false negatives (Type II errors) while tolerating some false positives (Type I errors) and redundant efforts. The checklist is organized sequentially, guiding auditors from a general understanding of the contract to a deeper analysis of vulnerabilities and best practices.

The checklist is divided into the following sections: \textbf{Contract Design}, \textbf{Asynchronous Execution}, \textbf{Common Vulnerabilities}, \textbf{Gas Control}, \textbf{Random Number Generation}, \textbf{Language-Specific Errors (FunC and Tact)}, and \textbf{Best Practices}. Below we provide an overview of each section with key details.

\subsection{Contract Design}
Start with mapping all message flows to understand how messages are processed and routed. Identify all entry points and analyze how input data and incoming messages are handled, including error handling. Ensure that the contract has robust authorization checks for all functions and message handlers. Evaluate the contract structure for unnecessary centralization and consider how the contract handles partial execution of transactions, especially in cases of gas exhaustion. Additionally, examine mechanisms to prevent the contract from being frozen or deleted.

\subsection{Asynchronous Execution}
TON's asynchronous execution model introduces unique challenges. Messages are guaranteed to be delivered, but not in a predictable timeframe, and their order is only predictable if sent to the same contract (using \textbf{logical time}). Consider the impact of concurrent processes and how state changes might occur between message flows. Analyze external dependencies and ensure that operations are independent of the sequence of message arrivals. Special attention should be paid to delayed state updates and race conditions, which are more prevalent in asynchronous systems.

\subsection{Common Vulnerabilities}
Address the public nature of blockchain data, ensuring that sensitive information like passwords or keys is never transmitted. Properly handle bounced messages to avoid issues such as tokens being sent to the void. Guard against unrestricted data recording, logical errors, and replay attacks by using mechanisms like \textbf{seqno}. Implement the \textbf{carry-value pattern} for inter-contract value transfers and ensure storage compatibility during smart contract updates. Avoid mixing signed and unsigned numbers in calculations and use exit codes correctly, avoiding reserved values. Prevent denial-of-service attacks by avoiding infinite or excessively long loops and ensure proper parsing and serialization of data.

\subsection{Gas Control}
Carefully calculate gas costs to ensure sufficient gas for operations and implement logic to return excess gas to the sender. Avoid data structures that can grow infinitely, as they increase gas costs over time. Recognize the risks of partial fulfillment when running out of gas and optimize storage parsing efficiency to minimize gas consumption. Additionally, be cautious when using modes like \texttt{SendRemainingValue} or \texttt{mode=64/128}, as they can deplete contract balances and cause subsequent operations to fail.

\subsection{Random Number Generation}
Be cautious with randomness generation, as validators can influence the outcome. In FunC, ensure \texttt{randomize\_lt()} or \texttt{randomize(x)} is used with \texttt{random()}. In Tact, use \textbf{nativeRandom} or \textbf{nativeRandomInterval} instead of \textbf{randomInt} and \textbf{random} to ensure secure randomness. Avoid predictable randomness formulas that attackers could exploit.

\subsection{Language-Specific Errors}
In FunC, use the \textbf{impure} modifier correctly for state-changing functions and carefully manage storage, variable ordering, and variable overrides. Avoid redeclaring variables to prevent namespace contamination. In Tact, avoid modifying parameters of incoming messages or functions and exercise caution when using assembler and native functions. Ensure proper initialization of variables in \texttt{init()} and avoid direct modification of inherited variables from traits.

\subsection{Best Practices}
Replace magic numbers with named constants for clarity and maintain thorough documentation of the contract's functionality and design decisions. Conduct independent code reviews and ensure that the contract adheres to relevant TON standards and best practices. Additionally, implement mechanisms to handle edge cases and unexpected behaviors, such as partial execution and external message handling.

\subsection{Checklist in Numbers}
By following this checklist, auditors can systematically assess the security and robustness of TON smart contracts, identifying potential vulnerabilities and ensuring reliable operation within the TON ecosystem. Figure \ref{fig:vulnerabilities_tree} provides a breakdown of vulnerabilities identified in TON smart contract audits, categorized according to the checklist. It is important to note that some vulnerabilities can be characterized by multiple categories and the categorization reflects the judgment of labeling specialists, which may vary depending on interpretation.

Logical errors dominate the \textbf{Common Errors} category, accounting for 70 out of 92 vulnerabilities. These errors, which include flaws in formulas, algorithms, and state transitions, highlight the critical need for rigorous testing and validation of contract logic. However, some of these logical errors could also overlap with the \textbf{Contract Design} category, particularly when they involve flawed business logic or state management. In the \textbf{Best Practices} category, documentation-related issues make up 23 out of 36 vulnerabilities, underscoring the importance of clear and thorough documentation to ensure that contract functionality and design decisions are well-understood and auditable. Documentation issues, while distinct, can also influence the clarity of other categories, such as authorization checks or gas control.

Authorization checks represent a significant portion of the \textbf{Contract Design} category, with 25 out of 76 vulnerabilities. This emphasizes the necessity of robust access control mechanisms to prevent unauthorized function calls and privilege escalation. However, access control issues can also intersect with the \textbf{Common Errors} category, especially when they involve logical flaws in permission checks. Meanwhile, the \textbf{Gas Control} category reveals 18 vulnerabilities related to inefficient gas handling, including failures to return excess gas and risks of partial fulfillment due to gas exhaustion. These issues can lead to significant operational challenges if not addressed and may overlap with the \textbf{Asynchronous Execution} category, where gas exhaustion can exacerbate race conditions or delayed state updates.

The \textbf{Asynchronous Execution} category highlights 6 vulnerabilities, all related to key considerations such as message delivery guarantees and state changes between messages. This reflects the unique challenges posed by TON's asynchronous model, particularly in high-frequency transaction environments where race conditions and delayed state updates can have severe consequences. Some of these vulnerabilities could also be classified under \textbf{Common Errors} or \textbf{Contract Design}, depending on the specific context and interpretation.

Interestingly, the \textbf{Random Number Generation} category includes only 1 vulnerability, but it is an important one. Secure randomness generation is essential to prevent manipulation by validators or attackers, especially in applications like gaming and lotteries. This vulnerability could also be linked to logical errors or design flaws, depending on the implementation. In contrast, the \textbf{Possible Errors in FunC} category is relatively minimal, with only 4 vulnerabilities identified. Modifying variables is the most prevalent issue within this category, with only two entries identified. This suggests that FunC's design is relatively robust in this regard. However, careful attention remains essential for state management and the proper use of function modifiers. These issues could also overlap with the \textbf{Contract Design} category, particularly when they involve storage management or variable ordering.

These facts highlight the areas where TON smart contracts are most vulnerable and provide actionable insights for auditors and developers. However, the categorization of vulnerabilities is not always clear-cut and depends on the judgment of labeling specialists. By systematically addressing these issues, the security and reliability of TON-based projects can be significantly improved.

\begin{figure}[htbp]
\centering
\begin{forest}
for tree={
  grow'=0,
  child anchor=west,
  parent anchor=east,
  anchor=west,
  calign=first,
  edge path={
    \noexpand\path[\forestoption{edge}]
    (!u.south west) +(5pt,0) |- (.child anchor)\forestoption{edge label};
  },
  before typesetting nodes={
    if n=1
      {insert before={[,phantom]}}
      {}
  },
  fit=band,
  before computing xy={l=12pt},
  s sep=2pt,
  l sep=8pt,
}
[Vulnerabilities 233
  [Contract Design 76
    [Authorization Checks 25]
    [Contract Design and Centralization 19]
    [External Message Handling 1]
    [Input Data Processing 15]
    [Logical Errors 8]
    [Message Generation and Handling 2]
    [Partial Execution of Transactions 6]
  ]
  [Asynchronous Execution 6
    [Key Considerations 6]
  ]
  [Common Errors 92
    [Bounced Message Handlers 3]
    [Carry-Value Pattern 1]
    [Exit Codes 2]
    [Logical Errors 70]
    [Parsing and Serialization 6]
    [Public Nature of Blockchain 2]
    [Replay Attack 3]
    [Restrictions on Data Recording 1]
    [Sending Messages from Loops 1]
    [Smart Contract Update 2]
    [Smart Contract Updates 1]
  ]
  [Gas Control 18
    [Moderate Handling of Gas 18]
  ]
  [Random Number Generation in TON 1
    [Safe Randomness Generation 1]
  ]
  [Possible Errors in FunC 4
    [Function Modifiers 1]
    [Modifying Variables 2]
    [Storage Management 1]
  ]
  [Best Practices 36
    [Code Review 7]
    [Compliance with Standards 5]
    [Documentation 23]
    [Magic Numbers-Flags-and Constants 1]
  ]
]
\end{forest}
\caption{The number of vulnerabilities in TON smart contracts audits per checklist category.}
\label{fig:vulnerabilities_tree}
\end{figure}

\section{Vulnerabilities Case Study}
\label{section:VulnerabilitiesCaseStudy}

In this section, we delve into specific vulnerabilities observed in TON smart contracts, focusing on challenges introduced by asynchronous execution, contract design flaws, and other critical issues. We present case studies from real-world projects highlighting the implications of these vulnerabilities and lessons learned.

\subsection{Contract Design}
Flaws in contract design often result in functional inconsistencies and financial losses. We examine a case from Onton Finance \cite{Beosin2024Onton} to illustrate this.

\subsubsection{Onton Finance: Lack of Rebound Mechanism}
In Onton Finance, the \texttt{Pool\_account} contract temporarily stores user assets during liquidity provision. The \texttt{withdraw\_liquidity} operation sets the ledger to zero before confirming the withdrawal. If the withdrawal fails, the user's assets are lost due to the lack of a rebound mechanism. This vulnerability underscores the importance of robust error handling in contract design.

\subsection{Asynchronous Execution}
TON's asynchronous execution model, while powerful, introduces unique challenges. Messages are guaranteed to be delivered, but not in a predictable timeframe, leading to potential race conditions and state inconsistencies. Below, we examine three cases from TONCO \cite{Beosin2024TONCO} and Storm Trade \cite{Quantstamp2024Storm} that illustrate these risks.

\subsubsection{TONCO: Incentive Calculation Flaws}
In TONCO, liquidity providers earn rewards based on the fee growth state of the liquidity pool. Delays in updating the \texttt{PositionNFT} contract can lead to inaccurate reward calculations. Consider the following sequence:

\begin{enumerate}
    \item At time \( t_0 \), Alice provides liquidity. The pool state is recorded as 100 in the \texttt{PositionNFT} contract.
    \item At time \( t_1 \), the pool-wide fee growth increases to 110 due to trading activity, but Alice's \texttt{PositionNFT} contract remains at 100.
    \item At time \( t_2 \), Alice removes liquidity. The \texttt{PoolContract} updates first, reflecting the correct fee growth (110), but the \texttt{PositionNFT} contract fails to update in time, returning an outdated or uninitialized value (0).
\end{enumerate}

This discrepancy results in an incorrect reward calculation: \( 0 - 110 = -110 \) instead of \( 110 - 100 = 10 \). The delayed state update highlights the risks of asynchronous execution in high-frequency transaction environments.

\subsubsection{Storm Trade: Race Conditions in Reward Distribution}
In Storm Trade, the \texttt{Executor} and \texttt{Referral Item} contracts manage the reward distribution. Asynchronous execution introduces race conditions, leading to incorrect balance updates. Two cases illustrate this:

\begin{enumerate}
    \item \textbf{Case 1: Overwriting Rewards} \\
    Bob makes two trades, triggering two referral rewards for Alice. The vault sends two messages to the \texttt{Referral Item} contract to add 5 TON each. Both messages independently verify the vault's whitelist status with the \texttt{Collection} contract. The second response overwrites the first, causing Alice's balance to remain 5 TON instead of 10 TON.

    \item \textbf{Case 2: Incorrect Balance Restoration} \\
    Alice requests a full withdrawal of 100 TON. The contract sets her balance to 0 and sends a withdrawal request to the vault. During this process, a reward update adds 10 TON to her balance. When the withdrawal fails, the bounce message restores Alice's balance to 100 TON, ignoring the 10 TON reward update.
\end{enumerate}

These cases demonstrate how race conditions in asynchronous systems can lead to lost rewards and incorrect balances.

\subsection{Common Errors}
Logical errors in smart contracts can lead to significant functional issues. We examine two cases: ThunderFinance \cite{Tonbit2024Thunder} and EVAA \cite{Quantstamp2024Evaa}.

\subsubsection{ThunderFinance: Incorrect LP Supply Adjustment}
In ThunderFinance, the \texttt{lpSupply} value increases during deposits but does not decrease during withdrawals. This flaw causes \texttt{lpSupply} to grow indefinitely, leading to inaccurate reward calculations and diminishing user rewards over time.

\subsubsection{EVAA: Incorrect Withdrawal Amount Calculation}
In EVAA, \texttt{calculate\_maximum\_withdraw\_amount()} function fails to account for varying collateral factors, resulting in over-collateralization and potential liquidation for users.

\subsection{Gas Control}
Improper gas handling can disrupt contract operations. We examine cases from ThunderFinance \cite{Tonbit2024Thunder} and TOM PUMP \cite{TonBit2024TOMPUMP}.

\subsubsection{ThunderFinance: Uncalculated Gas and Unprocessed Bounce}
In ThunderFinance, contracts fail to calculate gas consumption during operations like deposits and withdrawals. This oversight leads to state inconsistencies and a potential loss of user funds.

\subsubsection{TOM PUMP: Fee Omission in Swap Operations}
In TOM PUMP, the swap operation does not charge gas fees when slippage protection is triggered. This flaw results in financial losses for the protocol.

\subsection{Random Number Generation}
TON's randomness generation is susceptible to manipulation if not implemented securely. We analyze a case from the TRS 404 Smart Contract \cite{Tonbit2024TRC404}.

\subsubsection{TRS 404: Manipulable NFT Level}
The \texttt{deployNftItem} function in TRS 404 uses the \texttt{random()} function to generate pseudo-random numbers. These numbers are predictable, allowing attackers to manipulate with NFT levels. A secure solution involves using \texttt{randomize\_lt()} to incorporate the logical time into randomness generation, ensuring different results for each transaction.

\subsection{Possible Errors in FunC}
FunC-specific errors can lead to unintended contract behavior. We analyze a case from \textbf{EVAA} \cite{Quantstamp2024Evaa}.

\subsubsection{EVAA: Missing \texttt{impure} Specifier}
In EVAA, certain functions (e.g., \texttt{ton::cell\_fwd\_fee()}) can revert but lack the \texttt{impure} specifier. This oversight allows the FunC compiler to delete these function calls, leading to incorrect program execution.

\subsection{Best Practices}
To ensure the security and robustness of TON smart contracts, developers should adhere to best practices. We examine a case from \textbf{Grishmans Kombat} \cite{HashEx2024Grishmans}.  

\subsubsection{Grishmans Kombat: Testing Gaps}  
Grishmans Kombat lacked comprehensive tests and relied on outdated token interaction scripts incompatible with the current vault version. This absence of testing poses risks, as smart contracts are immutable post-deployment. Implement unit tests and validate on testnets to ensure functionality and security.

\section{Survey Validation of the Audit Checklist}
\label{subsection:SurveyValidation}
To validate the practical utility of our audit checklist, we conducted a survey~\cite{PositiveSecurity2025} with experienced TON ecosystem participants. We received 12 responses in total; one was excluded from quantitative analysis due to incompleteness (missing $>$40\% of items), leaving 11 complete responses for analysis. Respondents had high familiarity with TON (mean rating: 4.5/5) and substantial audit experience (8 of 11 had audited TON contracts, with 6 having audited more than 5 contracts).

\subsection{Quantitative Feedback}
Respondents rated the checklist highly across multiple dimensions (Table~\ref{tab:survey_ratings}). Notably, the checklist received strong scores for providing value beyond automated tools alone (mean: 4.4/5) and for minimal time overhead during audits (mean: 4.2/5). The category structure was rated as reflecting real-world audit priorities (mean: 4.1/5).

\begin{table}[h!]
\caption{Survey Ratings of Checklist Characteristics (1-5 scale, n=11)}
\label{tab:survey_ratings}
\centering
\begin{tabular}{lc}
\toprule
\textbf{Characteristic} & \textbf{Mean Rating} \\
\midrule
Covers full vulnerability range & 3.9 \\
Categories reflect audit priorities & 4.1 \\
Complements automated tools & 4.4 \\
Easy to use during audit & 4.1 \\
Logical structure & 3.8 \\
Minimal time overhead & 4.2 \\
Would integrate into workflow & 4.0 \\
Would recommend to peers & 3.9 \\
\bottomrule
\end{tabular}
\end{table}

\subsection{Qualitative Insights}
Respondents highlighted several strengths regarding the checklist's practical grounding:
\begin{itemize}
    \item \textit{Practical grounding}: One auditor noted that ``This TON checklist feels more practically grounded in how TON projects actually break down. It covers asynchronous execution issues and other TON-specific vulnerabilities that still often require manual reasoning rather than being caught by existing tooling.''
    \item \textit{Comparison to Ethereum}: A developer compared it to existing resources: ``Compared to Ethereum checklists (ConsenSys, SWC Registry), this checklist is well-structured and covers key TON-specific areas effectively: async execution, carry-value pattern, gas control, FunC/Tact language pitfalls.''
    \item \textit{Hybrid workflow}: Most respondents (6 of 11) preferred using the checklist alongside automated tools (e.g., Misti, TSA), indicating complementary rather than redundant value.
\end{itemize}

\subsection{Areas for Enhancement}
Feedback identified opportunities for checklist refinement, including requests for Tolk-specific guidance, explicit coverage of TVM limits (\texttt{max\_vm\_data\_depth}), and advanced patterns like cross-contract asynchronous reentrancy. These insights confirm the checklist's relevance while guiding iterative improvements for future releases.

\section{Discussion}
\label{section:Discussion}
False negatives in smart contract auditing, where vulnerabilities remain undetected, pose significant risks in the TON blockchain ecosystem. Such oversights can lead to financial losses, reputational damage, and exploitation by malicious actors. In TON's asynchronous systems, undetected race conditions or delayed state updates can result in inconsistent contract states or incorrect reward distributions, particularly in high-stakes areas like decentralized finance (DeFi) and gaming. Our audit checklist aims to minimize these risks by systematically addressing TON-specific vulnerabilities, such as message ordering issues and logical errors, while accepting some false positives to ensure comprehensive coverage.

Our work complements the recent paper \cite{Song2025} by Song et al., which focuses on automated defect detection in TON smart contracts using the static analysis tool TONScanner. Whilst Song et al. emphasize community-developed contracts and automated methods, our paper targets professional audit reports, offering a structured, heuristic checklist for developers and auditors. We provide practical, actionable insights and onboarding materials, with a particular focus on TON's asynchronous execution model, a topic not explicitly covered by Song et al. Together, these studies form a robust foundation for TON smart contract security, with our checklist serving as a manual guide and TONScanner providing automated support.

\subsection{Broader Implications for Asynchronous Systems}
While focused on TON, the challenges identified here resonate with broader software engineering contexts. TON's asynchronous message-passing model mirrors challenges found in event-driven architectures and microservices, where race conditions and state consistency across distributed components are critical concerns. The checklist patterns developed for TON--such as the carry-value pattern for atomic transfers and explicit state validation before message processing--offer transferable insights for securing any asynchronous distributed system where operation ordering cannot be guaranteed.

\subsection{Limitations}
Several limitations of this study warrant mention. First, our survey sample, while composed of experienced practitioners, remains limited in size (11 complete responses out of 12 received) and may not represent the full diversity of TON development practices. We excluded one partial response to maintain data integrity for quantitative metrics. Second, the checklist ratings reflect self-reported perceptions rather than controlled empirical evaluation of defect detection effectiveness. Despite these constraints, the strong practitioner endorsement (mean intent-to-use rating: 4.0/5) suggests practical relevance. Future work should include larger-scale validation across diverse TON projects and measurement of actual defect detection rates when the checklist is applied in real audit engagements.

The implications of this work are significant for the TON ecosystem. By adopting our checklist, developers and auditors can systematically identify and mitigate vulnerabilities, reducing the risk of exploits. Additionally, our findings underscore the importance of community collaboration in refining security practices, ensuring the long-term success of the TON ecosystem. 

\section{Conclusion}  
\label{section:Conclusion}  

Our analysis highlights the unique challenges posed by TON’s asynchronous execution model, such as delayed state updates and race conditions, which require careful consideration during development and auditing. Logical errors, particularly in state transitions and reward calculations, are the most common vulnerabilities in TON smart contracts, while improper gas handling can lead to partial transaction execution and financial losses.  

This paper contributes to the field by presenting a comprehensive checklist for auditing TON smart contracts, addressing vulnerabilities specific to TON’s architecture. It also offers detailed case studies of real-world vulnerabilities, providing practical insights and lessons for developers and auditors. Additionally, our checklist serves as a valuable onboarding resource for new participants in the TON ecosystem.  

Adopting this checklist can help to systematically identify and mitigate vulnerabilities, reducing the risk of exploits. Our work also encourages community collaboration to refine security practices and develop specialized tools. Future research should focus on developing automated detection tools, creating a formal grammar for the TON Virtual Machine (TVM), and extending the study to analyze complex contract interactions and multi-contract systems.  

Proactive security measures are essential for the sustainable growth of the TON ecosystem. By adopting rigorous audit practices and leveraging tools like the checklist presented in this paper, developers and auditors can enhance the security and reliability of TON smart contracts. We encourage the TON community to collaborate by refining these practices and developing innovative solutions to address emerging challenges.

\section*{Acknowledgment}

The research has been supported by the Ministry of Economic Development of the Russian Federation (agreement with MIPT No. 139-15-2025-013, dated June 20, 2025, IGK 000000C313925P4B0002).

We acknowledge the use of ChatGPT and DeepSeek in enhancing the readability and clarity of this manuscript. These tools were employed to assist in refining language and improving the overall presentation of the content. However, the authors retain full responsibility for the integrity, accuracy, and intellectual contributions of the research at all stages.

\bibliographystyle{elsarticle-num}
\bibliography{referencesLocal}

\begin{thebibliography}{10}
\expandafter\ifx\csname url\endcsname\relax
  \def\url#1{\texttt{#1}}\fi
\expandafter\ifx\csname urlprefix\endcsname\relax\def\urlprefix{URL }\fi
\expandafter\ifx\csname href\endcsname\relax
  \def\href#1#2{#2} \def\path#1{#1}\fi

\bibitem{Durov2019}
N.~Durov, \href{https://test.ton.org/tblkch.pdf}{{Telegram Open Network}} (2019) 1--132.
\newline\urlprefix\url{https://test.ton.org/tblkch.pdf}

\bibitem{Berger2019}
M.~P. Berger, J.~G. Tenreiro, K.~McGrath, \href{https://www.sec.gov/files/litigation/complaints/2019/comp-pr2019-212.pdf}{{SEC against Telegram Group inc. and TON issuer inc.}}, Tech. rep., SECURITIES AND EXCHANGE COMMISSION (2019).
\newline\urlprefix\url{https://www.sec.gov/files/litigation/complaints/2019/comp-pr2019-212.pdf}

\bibitem{TON2025}
{TON}, \href{https://ton.org/toncoin}{{Toncoin: The future of currency}}.
\newline\urlprefix\url{https://ton.org/toncoin}

\bibitem{TON2025a}
{TON}, \href{https://docs.ton.org/}{{Welcome to the TON Blockchain documentation}}.
\newline\urlprefix\url{https://docs.ton.org/}

\bibitem{Durov2020}
N.~Durov, \href{https://ton-blockchain.github.io/docs/tvm.pdf}{{Telegram Open Network Virtual Machine}}, Tech. rep. (2020).
\newline\urlprefix\url{https://ton-blockchain.github.io/docs/tvm.pdf}

\bibitem{Buterin2014}
V.~Buterin, \href{https://github.com/ethereum/wiki/wiki/White-Paper}{{Ethereum White Paper: A Next Generation Smart Contract {\&} Decentralized Application Platform}}, Ethereum~(January) (2014) 1--36.
\newline\urlprefix\url{https://github.com/ethereum/wiki/wiki/White-Paper}

\bibitem{Atzei2017}
N.~Atzei, M.~Bartoletti, T.~Cimoli, \href{http://link.springer.com/10.1007/978-3-662-54455-6_8}{{A Survey of Attacks on Ethereum Smart Contracts (SoK)}}, in: Lecture Notes in Computer Science (including subseries Lecture Notes in Artificial Intelligence and Lecture Notes in Bioinformatics), Vol. 10204 LNCS, Springer Verlag, 2017, pp. 164--186.
\newblock \href {https://doi.org/10.1007/978-3-662-54455-6_8} {\path{doi:10.1007/978-3-662-54455-6_8}}.
\newline\urlprefix\url{http://link.springer.com/10.1007/978-3-662-54455-6_8}

\bibitem{Chen2022a}
J.~Chen, X.~Xia, D.~Lo, J.~Grundy, X.~Luo, T.~Chen, \href{https://ieeexplore.ieee.org/document/9072659/}{{Defining Smart Contract Defects on Ethereum}}, IEEE Transactions on Software Engineering 48~(1) (2022) 327--345.
\newblock \href {https://doi.org/10.1109/TSE.2020.2989002} {\path{doi:10.1109/TSE.2020.2989002}}.
\newline\urlprefix\url{https://ieeexplore.ieee.org/document/9072659/}

\bibitem{Zhang2020b}
P.~Zhang, F.~Xiao, X.~Luo, {A Framework and DataSet for Bugs in Ethereum Smart Contracts}, in: 2020 IEEE International Conference on Software Maintenance and Evolution (ICSME), IEEE, 2020, pp. 139--150.
\newblock \href {https://doi.org/10.1109/ICSME46990.2020.00023} {\path{doi:10.1109/ICSME46990.2020.00023}}.

\bibitem{Soud2024}
M.~Soud, G.~Liebel, M.~Hamdaqa, {A fly in the ointment: an empirical study on the characteristics of Ethereum smart contract code weaknesses}, Empirical Software Engineering 29~(1) (2024) 13.
\newblock \href {https://doi.org/10.1007/s10664-023-10398-5} {\path{doi:10.1007/s10664-023-10398-5}}.

\bibitem{He2020}
D.~He, Z.~Deng, Y.~Zhang, S.~Chan, Y.~Cheng, N.~Guizani, {Smart Contract Vulnerability Analysis and Security Audit}, IEEE Network 34~(5) (2020) 276--282.
\newblock \href {https://doi.org/10.1109/MNET.001.1900656} {\path{doi:10.1109/MNET.001.1900656}}.

\bibitem{Khan2020}
Z.~A. Khan, A.~Siami~Namin, {Ethereum Smart Contracts: Vulnerabilities and their Classifications}, in: 2020 IEEE International Conference on Big Data (Big Data), IEEE, 2020, pp. 1--10.
\newblock \href {https://doi.org/10.1109/BigData50022.2020.9439088} {\path{doi:10.1109/BigData50022.2020.9439088}}.

\bibitem{Khan2024}
Z.~A. Khan, A.~S. Namin, {A Survey of Vulnerability Detection Techniques by Smart Contract Tools}, IEEE Access 12 (2024) 70870--70910.
\newblock \href {https://doi.org/10.1109/ACCESS.2024.3401623} {\path{doi:10.1109/ACCESS.2024.3401623}}.

\bibitem{Feist2019}
J.~Feist, G.~Grieco, A.~Groce, {Slither: A static analysis framework for smart contracts}, in: Proceedings - 2019 IEEE/ACM 2nd International Workshop on Emerging Trends in Software Engineering for Blockchain, WETSEB 2019, Institute of Electrical and Electronics Engineers Inc., 2019, pp. 8--15.
\newblock \href {https://doi.org/10.1109/WETSEB.2019.00008} {\path{doi:10.1109/WETSEB.2019.00008}}.

\bibitem{Consensys2025}
{Consensys}, \href{https://mythx.io/}{{MythX: Smart contract security service for Ethereum}} (2025).
\newline\urlprefix\url{https://mythx.io/}

\bibitem{Consensys2025a}
{Consensys}, \href{https://github.com/ConsenSys/mythril}{{Mythril: symbolic-execution-based securty analysis tool for EVM bytecode. It detects security vulnerabilities in smart contracts built for Ethereum and other EVM-compatible blockchains}} (2025).
\newline\urlprefix\url{https://github.com/ConsenSys/mythril}

\bibitem{Wustholz2020}
V.~W{\"{u}}stholz, M.~Christakis, {Harvey: a greybox fuzzer for smart contracts}, in: Proceedings of the 28th ACM Joint Meeting on European Software Engineering Conference and Symposium on the Foundations of Software Engineering, ACM, New York, NY, USA, 2020, pp. 1398--1409.
\newblock \href {https://doi.org/10.1145/3368089.3417064} {\path{doi:10.1145/3368089.3417064}}.

\bibitem{deMoura2008}
L.~de~Moura, N.~Bj{\o}rner, {Z3: An Efficient SMT Solver}, 2008, pp. 337--340.
\newblock \href {https://doi.org/10.1007/978-3-540-78800-3_24} {\path{doi:10.1007/978-3-540-78800-3_24}}.

\bibitem{Grieco2020}
G.~Grieco, W.~Song, A.~Cygan, J.~Feist, A.~Groce, {Echidna: effective, usable, and fast fuzzing for smart contracts}, in: Proceedings of the 29th ACM SIGSOFT International Symposium on Software Testing and Analysis, ACM, New York, NY, USA, 2020, pp. 557--560.
\newblock \href {https://doi.org/10.1145/3395363.3404366} {\path{doi:10.1145/3395363.3404366}}.

\bibitem{Nguyen2020a}
T.~D. Nguyen, L.~H. Pham, J.~Sun, Y.~Lin, Q.~T. Minh, {sFuzz: an efficient adaptive fuzzer for solidity smart contracts}, in: Proceedings of the ACM/IEEE 42nd International Conference on Software Engineering, ACM, New York, NY, USA, 2020, pp. 778--788.
\newblock \href {https://doi.org/10.1145/3377811.3380334} {\path{doi:10.1145/3377811.3380334}}.

\bibitem{Kalra2018}
S.~Kalra, S.~Goel, M.~Dhawan, S.~Sharma, \href{https://www.ndss-symposium.org/wp-content/uploads/2018/02/ndss2018_09-1_Kalra_paper.pdf}{{ZEUS: Analyzing Safety of Smart Contracts}}, in: Proceedings 2018 Network and Distributed System Security Symposium, Internet Society, Reston, VA, 2018.
\newblock \href {https://doi.org/10.14722/ndss.2018.23082} {\path{doi:10.14722/ndss.2018.23082}}.
\newline\urlprefix\url{https://www.ndss-symposium.org/wp-content/uploads/2018/02/ndss2018_09-1_Kalra_paper.pdf}

\bibitem{Hildenbrandt2018}
E.~Hildenbrandt, M.~Saxena, N.~Rodrigues, X.~Zhu, P.~Daian, D.~Guth, B.~Moore, D.~Park, Y.~Zhang, A.~Stefanescu, G.~Rosu, \href{http://kframework.org/}{{KEVM: A complete formal semantics of the ethereum virtual machine}}, in: Proceedings - IEEE Computer Security Foundations Symposium, Vol. 2018-July, 2018, pp. 204--217.
\newblock \href {https://doi.org/10.1109/CSF.2018.00022} {\path{doi:10.1109/CSF.2018.00022}}.
\newline\urlprefix\url{http://kframework.org/}

\bibitem{Certora2025}
{Certora}, \href{https://www.certora.com/prover}{{Certora Prover}}.
\newline\urlprefix\url{https://www.certora.com/prover}

\bibitem{Annenkov2020}
D.~Annenkov, J.~B. Nielsen, B.~Spitters, {ConCert: a smart contract certification framework in Coq}, in: Proceedings of the 9th ACM SIGPLAN International Conference on Certified Programs and Proofs, ACM, New York, NY, USA, 2020, pp. 215--228.
\newblock \href {https://doi.org/10.1145/3372885.3373829} {\path{doi:10.1145/3372885.3373829}}.

\bibitem{Kevin2025}
J.~Kevin, P.~Yugopuspito, {SmartLLM: Smart Contract Auditing using Custom Generative AI}, Arxiv (2 2025).

\bibitem{Ma2025}
W.~Ma, D.~Wu, Y.~Sun, T.~Wang, S.~Liu, J.~Zhang, Y.~Xue, Y.~Liu, {Combining Fine-Tuning and LLM-based Agents for Intuitive Smart Contract Auditing with Justifications}, in: 2025 IEEE/ACM 47th International Conference on Software Engineering (ICSE), 2025, pp. 330--342.
\newblock \href {https://doi.org/10.1109/ICSE55347.2025.00027} {\path{doi:10.1109/ICSE55347.2025.00027}}.

\bibitem{Song2025}
H.~Song, T.~Li, J.~Chen, T.~Chen, B.~Li, Z.~Lin, Y.~Lu, P.~Li, X.~Zhou, {Enhancing The Open Network: Definition and Automated Detection of Smart Contract Defects}, Arxiv (1 2025).

\bibitem{Espirito2025}
{Espirito}, \href{https://github.com/espritoxyz/tsa}{{TSA: TON Symbolic Analyzer}} (2025).
\newline\urlprefix\url{https://github.com/espritoxyz/tsa}

\bibitem{Tonbit2023TonUp}
{TonBit}, \href{https://tonbit.xyz/reports/TonUP-Smart-Contract-Final-Audit-Report.pdf}{{TonUP Audit Report}}, Tech. rep. (2023).
\newline\urlprefix\url{https://tonbit.xyz/reports/TonUP-Smart-Contract-Final-Audit-Report.pdf}

\bibitem{Quantstamp2023TONLocker}
{Quantstamp}, \href{https://certificate.quantstamp.com/full/ton-locker-contract/6872997f-1110-45cc-b70f-2a4cd639da1f/index.html}{{TON Locker Contract Audit Report}}, Tech. rep. (2023).
\newline\urlprefix\url{https://certificate.quantstamp.com/full/ton-locker-contract/6872997f-1110-45cc-b70f-2a4cd639da1f/index.html}

\bibitem{Sedov2023Hipo}
D.~Sedov, \href{https://github.com/HipoFinance/audits/blob/main/Daniil%20Sedov%20Hipo%20Audit%20Report%202023-10.pdf}{{Hipo Finance Audit Report}}, Tech. rep. (2023).
\newline\urlprefix\url{https://github.com/HipoFinance/audits/blob/main/Daniil%20Sedov%20Hipo%20Audit%20Report%202023-10.pdf}

\bibitem{ScaleBit2023Hipo}
{ScaleBit}, \href{https://scalebit.xyz/reports/Hipo-Finance-Audit-Report.pdf}{{Hipo Finance Audit Report}}, Tech. rep. (2023).
\newline\urlprefix\url{https://scalebit.xyz/reports/Hipo-Finance-Audit-Report.pdf}

\bibitem{Beosin2024Aqua}
{Beosin}, \href{https://www.beosin.com/audits/Aqua%20Protocol_202407221416.pdf}{{Aqua Protocol Smart Contract Security Audit No.202407221416}}, Tech. rep. (2024).
\newline\urlprefix\url{https://www.beosin.com/audits/Aqua%20Protocol_202407221416.pdf}

\bibitem{Beosin2024InterBridge}
{Beosin}, \href{https://beosin.com/audits/InterBridge-Ton_202410161700.pdf}{{InterBridge-Ton Audit Report}}, Tech. rep. (2024).
\newline\urlprefix\url{https://beosin.com/audits/InterBridge-Ton_202410161700.pdf}

\bibitem{Beosin2024Onton}
{Beosin}, \href{https://beosin.com/audits/Onton_Finance_202409121334.pdf}{{Onton Finance Audit Report}}, Tech. rep. (2024).
\newline\urlprefix\url{https://beosin.com/audits/Onton_Finance_202409121334.pdf}

\bibitem{Beosin2024TONCO}
{Beosin}, \href{https://beosin.com/audits/TONCO_202411221000.pdf}{{TONCO Audit Report}}, Tech. rep. (2024).
\newline\urlprefix\url{https://beosin.com/audits/TONCO_202411221000.pdf}

\bibitem{Beosin2024Tonny}
{Beosin}, \href{https://beosin.com/audits/Tonny_202409231139.pdf}{{Tonny Audit Report}}, Tech. rep. (2024).
\newline\urlprefix\url{https://beosin.com/audits/Tonny_202409231139.pdf}

\bibitem{BugBlow2024Aqua}
{BugBlow}, \href{https://github.com/BugBlow/audits/blob/main/AquaProtocol/Aqua_Security_Audit_BugBlow.pdf}{{Aqua Protocol Security Audit}}, Tech. rep. (2024).
\newline\urlprefix\url{https://github.com/BugBlow/audits/blob/main/AquaProtocol/Aqua_Security_Audit_BugBlow.pdf}

\bibitem{BugBlow2024CryptoBillions}
{BugBlow}, \href{https://github.com/BugBlow/audits/blob/main/CryptoBillions/CryptoBillions_Audit_BugBlow.pdf}{{CryptoBillions Audit Report}}, Tech. rep. (2024).
\newline\urlprefix\url{https://github.com/BugBlow/audits/blob/main/CryptoBillions/CryptoBillions_Audit_BugBlow.pdf}

\bibitem{TonBit2024BeaverLand}
{TonBit}, \href{http://tonbit.xyz/reports/20241025-BeaverLand-Final-Audit-Report.pdf}{{BeaverLand Audit Report}}, Tech. rep. (2024).
\newline\urlprefix\url{http://tonbit.xyz/reports/20241025-BeaverLand-Final-Audit-Report.pdf}

\bibitem{TonBit2024Bool}
{TonBit}, \href{http://tonbit.xyz/reports/20241025-Bool-Network-Smart-Contract-Final-Audit-Report.pdf}{{Bool Network Audit Report}}, Tech. rep. (2024).
\newline\urlprefix\url{http://tonbit.xyz/reports/20241025-Bool-Network-Smart-Contract-Final-Audit-Report.pdf}

\bibitem{TonBit2024MiniTon}
{TonBit}, \href{http://tonbit.xyz/reports/20241025-Miniton-Smart-Contract-Final-Audit-Report.pdf}{{MiniTon Audit Report}}, Tech. rep. (2024).
\newline\urlprefix\url{http://tonbit.xyz/reports/20241025-Miniton-Smart-Contract-Final-Audit-Report.pdf}

\bibitem{TonBit2024OneClick}
{TonBit}, \href{http://tonbit.xyz/reports/20241025-One-Click-Sender-Final-Audit-Report.pdf}{{One Click Sender Audit Report}}, Tech. rep. (2024).
\newline\urlprefix\url{http://tonbit.xyz/reports/20241025-One-Click-Sender-Final-Audit-Report.pdf}

\bibitem{Tonbit2024SecondLive}
{TonBit}, \href{https://tonbit.xyz/reports/20240925-SecondLive-Ton-Final-Audit-Report.pdf}{{SecondLive-Ton Audit Report}}, Tech. rep. (2024).
\newline\urlprefix\url{https://tonbit.xyz/reports/20240925-SecondLive-Ton-Final-Audit-Report.pdf}

\bibitem{TonBit2024Catizen}
{TonBit}, \href{http://tonbit.xyz/reports/20240828-Catizen-Smart-Contarct-Final-Audit-Report.pdf}{{Catizen Audit Report}}, Tech. rep. (2024).
\newline\urlprefix\url{http://tonbit.xyz/reports/20240828-Catizen-Smart-Contarct-Final-Audit-Report.pdf}

\bibitem{TonBit2024CatizenJetton}
{TonBit}, \href{http://tonbit.xyz/reports/20240828-Catizen-Jetton-Smart-Contract-Final-Audit-Report.pdf}{{Catizen Jetton Audit Report}}, Tech. rep. (2024).
\newline\urlprefix\url{http://tonbit.xyz/reports/20240828-Catizen-Jetton-Smart-Contract-Final-Audit-Report.pdf}

\bibitem{Tonbit2024Thunder}
{TonBit}, \href{http://tonbit.xyz/reports/ThunderFinance-Final-Audit-Report.pdf}{{ThunderFinance Audit Report}}, Tech. rep. (2024).
\newline\urlprefix\url{http://tonbit.xyz/reports/ThunderFinance-Final-Audit-Report.pdf}

\bibitem{TonBit2024TOMPUMP}
{TonBit}, \href{https://tonbit.xyz/reports/20241023-TOM-PUMP-Final-Audit-Report.pdf}{{TOM PUMP Audit Report}}, Tech. rep. (2024).
\newline\urlprefix\url{https://tonbit.xyz/reports/20241023-TOM-PUMP-Final-Audit-Report.pdf}

\bibitem{Tonbit2024TonStaking}
{TonBit}, \href{https://tonbit.xyz/reports/20240930-Ton-Staking-Final-Audit-Report.pdf}{{Ton Staking Protocol Audit Report}}, Tech. rep. (2024).
\newline\urlprefix\url{https://tonbit.xyz/reports/20240930-Ton-Staking-Final-Audit-Report.pdf}

\bibitem{Tonbit2024Trapdoor}
{TonBit}, \href{https://www.tonbit.xyz/reports/Tradoor-Smart-Contract-Audit-Report-Summary.pdf}{{Tradoor Audit Report}}, Tech. rep. (2024).
\newline\urlprefix\url{https://www.tonbit.xyz/reports/Tradoor-Smart-Contract-Audit-Report-Summary.pdf}

\bibitem{Tonbit2024TRC404}
{TonBit}, \href{http://tonbit.xyz/reports/TRC404-Smart-Contract-Final-Audit-Report.pdf}{{TRC404 Audit Report}}, Tech. rep. (2024).
\newline\urlprefix\url{http://tonbit.xyz/reports/TRC404-Smart-Contract-Final-Audit-Report.pdf}

\bibitem{Tonbit2024UTonic}
{TonBit}, \href{https://tonbit.xyz/reports/20241012-UTonic-Final-Audit-Report.pdf}{{UTonic Audit Report}}, Tech. rep. (2024).
\newline\urlprefix\url{https://tonbit.xyz/reports/20241012-UTonic-Final-Audit-Report.pdf}

\bibitem{Quantstamp2024FDUSD}
{Quantstamp}, \href{https://certificate.quantstamp.com/full/fdusd-on-ton/8ce8359d-7f0e-476b-a4de-183cca98b8c8/index.html}{{FDUSD on TON Audit Report}}, Tech. rep. (2024).
\newline\urlprefix\url{https://certificate.quantstamp.com/full/fdusd-on-ton/8ce8359d-7f0e-476b-a4de-183cca98b8c8/index.html}

\bibitem{Quantstamp2024Storm}
{Quantstamp}, \href{https://certificate.quantstamp.com/full/storm-trade/21e4074a-b2cb-409d-b5df-48f683d0e8f3/index.html}{{Storm Trade Audit Report}}, Tech. rep. (2024).
\newline\urlprefix\url{https://certificate.quantstamp.com/full/storm-trade/21e4074a-b2cb-409d-b5df-48f683d0e8f3/index.html}

\bibitem{HashEx2024Grishmans}
{HashEx Blockchain Security}, \href{https://github.com/HashEx/public_audits/blob/master/Grishmans%20Kombat/Grishmans%20Kombat.pdf}{{Grishmans Kombat Audit Report}}, Tech. rep. (2024).
\newline\urlprefix\url{https://github.com/HashEx/public_audits/blob/master/Grishmans%20Kombat/Grishmans%20Kombat.pdf}

\bibitem{Quantstamp2024RhinoFi}
{Quantstamp}, \href{https://certificate.quantstamp.com/full/rhino-fi/6529d3d8-4906-43c9-bfe0-601ec83647cb/index.html}{{Security Assessment of Rhino Fi}}, Tech. rep. (2024).
\newline\urlprefix\url{https://certificate.quantstamp.com/full/rhino-fi/6529d3d8-4906-43c9-bfe0-601ec83647cb/index.html}

\bibitem{Quantstamp2024Evaa}
{Quantstamp}, \href{https://certificate.quantstamp.com/full/evaa/df7aa699-793b-49f7-b348-1f78e9ca9870/index.html}{{Security Assessment of Evaa}}, Tech. rep. (2024).
\newline\urlprefix\url{https://certificate.quantstamp.com/full/evaa/df7aa699-793b-49f7-b348-1f78e9ca9870/index.html}

\bibitem{TonTech2024Hipo}
{TonTech}, \href{https://github.com/HipoFinance/audits/blob/main/TonTech%20Hipo%20Audit%20Report%202023-10.pdf}{{Hipo Finance Audit Report}}, Tech. rep. (2024).
\newline\urlprefix\url{https://github.com/HipoFinance/audits/blob/main/TonTech%20Hipo%20Audit%20Report%202023-10.pdf}

\bibitem{ProgramCrafter2024Hipo}
{ProgramCrafter}, \href{https://github.com/HipoFinance/audits/blob/main/hTON/hTON_Audit_ProgramCrafter.pdf}{{hTON (Hipo Staking Protocol) Audit Report}}, Tech. rep. (2024).
\newline\urlprefix\url{https://github.com/HipoFinance/audits/blob/main/hTON/hTON_Audit_ProgramCrafter.pdf}

\bibitem{Chainsulting2024TONMS}
{Chainsulting}, \href{https://github.com/softstack/Smart-Contract-Security-Audits/blob/master/TON/Smart_Contract_Audit_TON_Multisig_18022023.pdf}{{TON Multisignature Wallet Audit Report}}, Tech. rep. (2024).
\newline\urlprefix\url{https://github.com/softstack/Smart-Contract-Security-Audits/blob/master/TON/Smart_Contract_Audit_TON_Multisig_18022023.pdf}

\bibitem{Softstackio2024XTON}
{Softstack.io}, \href{https://github.com/softstack/Smart-Contract-Security-Audits/blob/master/XTON/Smart_Contract_Audit_XTON_Core_21032024.pdf}{{XTON Core Audit Report}}, Tech. rep. (2024).
\newline\urlprefix\url{https://github.com/softstack/Smart-Contract-Security-Audits/blob/master/XTON/Smart_Contract_Audit_XTON_Core_21032024.pdf}

\bibitem{BugBlow2025Boxing}
{BugBlow}, \href{https://github.com/BugBlow/audits/blob/main/Delabs_TON_Security_Audit_Report_By_BugBlow.pdf}{{Boxing Star X Wallet Audit Report}}, Tech. rep. (2025).
\newline\urlprefix\url{https://github.com/BugBlow/audits/blob/main/Delabs_TON_Security_Audit_Report_By_BugBlow.pdf}

\bibitem{PositiveSecurity2025}
PositiveSecurity, \href{https://github.com/PositiveSecurity/ton-audit-guide/tree/paper}{Checklist for auditing ton smart contracts} (2025).
\newline\urlprefix\url{https://github.com/PositiveSecurity/ton-audit-guide/tree/paper}

\end{thebibliography}







\end{document}